\documentclass[aps,preprint,superscriptaddress,groupedaddress,nofootinbib]{revtex4}  

\usepackage{graphicx}  
\usepackage{dcolumn}   
\usepackage{bm}        
\usepackage{amssymb}   

\usepackage{color} 

\usepackage{amsmath}

\begin{document}

%
%

\title{Localization of Interacting Fields in Five-Dimensional Braneworld Models}

\author{Dewi Wulandari}
\thanks{wulandaridewi@unimed.ac.id}
\affiliation{Physics Department, Universitas Negeri Medan, Medan Estate 20221, Indonesia \\ and \\ Department of Physics, Faculty of Mathematics and Natural Sciences, Institut Teknologi Bandung, Jalan Ganesha 10 Bandung 40132, Indonesia}

\author{Triyanta}
\thanks{triyanta@fi.itb.ac.id}
\affiliation{Department of Physics, Faculty of Mathematics and Natural Sciences, Institut Teknologi Bandung, Jalan Ganesha 10 Bandung 40132, Indonesia}

\author{Jusak S. Kosasih}
\thanks{jusak@fi.itb.ac.id}
\affiliation{Department of Physics, Faculty of Mathematics and Natural Sciences, Institut Teknologi Bandung, Jalan Ganesha 10 Bandung 40132, Indonesia}

\author{Douglas Singleton}
\thanks{dougs@mail.fresnostate.edu}
\affiliation{California State University Fresno, Fresno, CA 93740}

\author{Preston Jones}
\thanks{Preston.Jones1@erau.edu}
\affiliation{Embry Riddle Aeronautical University, Prescott, AZ 86301}


\begin{abstract}
We study localization properties of fundamental fields which are coupled to one another through the gauge mechanism both in the original Randall-Sundrum (RS)  and in the modified Randall-Sundrum (MRS) braneworld models: scalar-vector, vector-vector, and spinor-vector configuration systems. For this purpose we derive conditions of localization, namely the finiteness of integrals over the extra coordinate in the action of the system considered. We also derive field equations for each of the systems and then obtain their solutions corresponding to the extra dimension by a separation of variable method for every field involved in each system. We then insert the obtained solutions into the conditions of localization to seek whether or not the solutions are in accordance with the conditions of localization. We obtain that not all of the configuration systems considered are localizable on the brane of the original RS model while, on the contrary, they are localizable on the MRS braneworld model with some restrictions. In terms of field localizability on the brane, this result shows that the MRS model is much better than the original RS model.

\end{abstract}

\maketitle


\section{Introduction}	

Extra dimensions were introduced in Kaluza-Klein models in order to unify electromagnetic and gravitational forces. The extra dimension is of the order of the Planck length so that direct experimental probing on it is hopeless \cite{Overduin}. Later on, extra dimensions were proposed to address several issues such as cosmological constant, dark matter, non-locality, and hierarchy problems \cite{Gogberashvili1, Gogberashvili2, Gogber2, Rubakov, Visser, Squires, Kalbermann, Gogberashvili3}. Here, matter fields and gauge interactions are assumed to be trapped in 4-dimensional sub-manifolds \cite{Rubakov, Visser,Squires, Barnaveli} commonly referred to as brane world models. The birth of braneworld models is also inspired by string theory  \cite{Polchi1, Horova}, a theory that was built to unify all interactions including gravity. 

The requirement of matter localization on the brane introduces various attempts to trap matter. An example is shown in Refs. \cite{Gogberashvili4, Midodashvili1, Midodashvili2}  that zero modes of all kinds of matter fields and four-gravity may be localized on the (1+3)-dimensional brane in a six-dimensional bulk by introducing an increasing and transverse gravitational potential. Other mechanisms to address this issue in 6D are given in Refs. \cite{Gogberashvili5, Midodashvili3, Midodashvili4}  and for higher dimensional brane world models in  Refs. \cite{Oda1, Oda2, Singleton}.

The RS model [22] which is characterized by a metric of the form
\begin{equation}
\label{eq1}
ds_{[y]} ^2 = a^{2}(x^{5})\eta_{\mu\nu}dx^{\mu}dx^{\nu}-dx^{5}dx^{5} ,
\end{equation}
with $a(x^5)=e^{-k|x^5|}$, $x^5=y$ is the fifth coordinate, $k$ is a constant and $\eta_{\mu\nu}$ is the Minkowski metric with signature $(1, -1, -1, -1)$,  is an appealing model as it resolves the hierarchy problem \cite{RaSu1}.  However,  this model is not a perfect braneworld model as not all types of fundamental matter fields are localized on 3+1 brane in a simple manner \cite {Gogber2,RaSu2}. In fact, only massless spin-0 fields are localized on the brane for decreasing warp factors \cite{Bajc}. For increasing warp factors, on the other hand, only spin-1/2 fields are localized on the brane \cite{Bajc}. Spin-1 fields are not localized for either a decreasing or an increasing warp factor \cite{Pomarol}. This fact led authors in Ref. \cite{Jones} to introduce the MRS model, specified by the metric 
\begin{equation}
\label{eq2}
ds_{[r]} ^2 = a^2(x^5)\eta_{\mu\nu}dx^\mu dx^\nu -b^2(x^5) dx^5dx^5 , b=a.
\end{equation}

We use $r$-coordinate instead of $y$-coordinate to indicate the fifth coordinate in the MRS model. The localization properties for the MRS model are better as compared to the RS model \cite {Jones, Triyanta}. This paper extends the work in Refs. \cite {Jones, Triyanta} by considering interacting fields, fundamental fields that interact with vector fields through a gauge mechanism in five-dimensional curved RS and MRS spacetimes, in addition to interacting with gravity.  In Section 3 we will look at the localization of interacting fields both in the RS and in the MRS models. Conclusions are given in Section 4. However, before proceeding, we first make some comments on Ref. \cite {Jones} regarding the localization of spin-0 and spin-1/2 fields in the MRS model.

\section{COMMENTS ON REF. \cite {Jones}}

The action of massless scalar fields that only interact with gravity  in a five-dimensional modified RS model is given by:
\begin{equation}
\label{eq3}
S  =  \int d^{5}x\sqrt{g}\partial_M\Phi^{*}\partial^M\Phi ,
\end{equation}
where $g$ is the determinant of the MRS metric and the capital Latin indices $M=0,1,2,3,5$. Decomposing $\Phi(x^M)=\varphi(x^\mu)\chi(x^5) $ the action becomes:
\begin{equation}
\label{eq4}
S  =  \int_0^\infty dr\sqrt{g} \chi^{*}\chi g^{\mu\nu}\int d^4x\partial_\mu\varphi^{*}\partial_\nu\varphi+ \int_0^\infty dr\sqrt{g}g^{rr}\partial_r\chi^{*} \partial_r\chi\int d^4x\varphi^*\varphi .
\end{equation}
The fields are said to be localized on the brane if the action integrals over the extra dimension from $0$ to $\infty$ are finite. It means that the conditions for localization are
\begin{subequations}
	\begin{equation}
	\int_0^\infty dr\sqrt{g} \chi^{*}\chi g^{\mu\nu}=\int_0^\infty dra^{2}b\chi^{*}\chi \eta^{\mu\nu}=N\eta^{\mu\nu},  \label{eq5a}
	\end{equation}
	\begin{equation}
	\int_0^\infty dr\sqrt{g}g^{rr}\partial_r\chi^{*} \partial_r\chi=-\int_0^\infty dr\frac{a^4}{b}\partial_r\chi^{*} \partial_r\chi=-m^2. \label{eq5b}
	\end{equation}
\end{subequations}
In the last equation of eq. (30) in Ref. \cite{Jones}, $\sqrt{g}g^{rr}$  is written as $-a^{2}/b$  which equals $-a$ in the MRS model since in this model, $b=a$. However this is incorrect because $\sqrt{g}=a^4b$, $g^{rr}=-1/b^{2}$  giving $\sqrt{g}g^{rr}=-a^{4}/b$ which is equivalent to $-a^3$ in the MRS model. The general conclusion is still correct that massless scalar fields are localizable on the brane for a decreasing warp factor while the massive ones are localizable both for a decreasing and an increasing warp factors.

Now we go to the case of spinor fields $\Psi$ in the five-dimensional MRS brane model. Ref. \cite{Jones} concludes that the spinor field is not localizable on the brane for both the decreasing and increasing warp factors. This result is based on the definition of adjoint of spinor fields that $\overline{\Psi}=\Psi^+\gamma^{\overline{0}}$ where $\gamma^{\overline{0}}$ is the zeroth  Dirac matric in four-dimensional Minkowski spacetime. However, if we define $\overline{\Psi}=\Psi^+\Gamma^0$ where $\Gamma^0$ is the zeroth  Dirac matric in the five-dimensional MRS curved spacetime the spinor field becomes localizable. The later definition of adjoint more makes sense since both $\Psi$ and $\Gamma^0$  are defined in five-dimensional curved spacetime. To prove the statement let us consider the following action
\begin{eqnarray}\label{eq6}
S=\int d^{5}x \sqrt{g}[\overline{\Psi}i\Gamma^MD_M\Psi],
\end{eqnarray} 
where \emph{g} is the determinant of the metric (\ref{eq2}), $\Gamma^M =e_{\overline{M}}^{M}\gamma^{\overline{M}}$ are gamma matrices in a curved spacetime, $\gamma^{\overline{M}} $ are  the gamma matrices in Minkowski spacetime with $\gamma^{\overline{5}}=i\gamma^{\overline{0}}\gamma^{\overline{1}}\gamma^{\overline{2}}\gamma^{\overline{3}}$ while $e_{M}^{\overline{M}} $ and $e_{\overline{M}}^{M}$ are funfbeins and their inverses, respectively, and $\overline{\Psi}=\Psi^+\Gamma^0=(\Psi^*)^Te_{\overline{0}}^{0}\gamma^{\overline{0}}$.  $D_M$ are the covariant derivatives defined in Ref. \cite{Jones}.

Decomposing the spinor  $\Psi(x^M)$ in a five dimensional spacetime as 
\begin{eqnarray}\label{eq7}
\Psi(x^{\mu},r)=\left(\begin{array}{c}\psi_R^{(1)}(x^\mu)P_R^{(1)}(r)\\\psi_R^{(2)}(x^\mu)P_R^{(2)}(r)\\\psi_L^{(1)}(x^\mu)P_L^{(1)}(r)\\\psi_L^{(2)}(x^\mu)P_L^{(2)}(r)\end{array}\right),
\end{eqnarray}
the action \eqref{eq6} becomes,
\begin{eqnarray}\label{eq8}
S&=&\int d^{5}x \frac{\sqrt{g}}{a^{2}(r)} [\psi_R^{(1)*}P_R^{(1)*}(i\partial_0\psi_R^{(1)})P_R^{(1)}+\psi_R^{(2)*}P_R^{(2)*}(i\partial_0\psi_R^{(2)})P_R^{(2)}\nonumber\\
&&+\psi_L^{(1)*}P_L^{(1)*}(i\partial_0\psi_L^{(1)})P_L^{(1)}+\psi_L^{(2)*}P_L^{(2)*}(i\partial_0\psi_L^{(2)})P_L^{(2)}] \nonumber\\
&&+\int d^{5}x \frac{\sqrt{g}}{a^{2}(r)} [\psi_R^{(2)*}P_R^{(2)*}(i\partial_1\psi_R^{(1)})P_R^{(1)}+\psi_R^{(1)*}P_R^{(1)*}(i\partial_1\psi_R^{(2)})P_R^{(2)}\nonumber\\
&&-\psi_L^{(2)*}P_L^{(2)*}(i\partial_1\psi_L^{(1)})P_L^{(1)}-\psi_L^{(1)*}P_L^{(1)*}(i\partial_1\psi_L^{(2)})P_L^{(2)}] \nonumber\\
&&+\int d^{5}x \frac{\sqrt{g}}{a^{2}(r)}i [\psi_R^{(2)*}P_R^{(2)*}(i\partial_2\psi_R^{(1)})P_R^{(1)}-\psi_R^{(1)*}P_R^{(1)*}(i\partial_2\psi_R^{(2)})P_R^{(2)}\nonumber\\
&&-\psi_L^{(2)*}P_L^{(2)*}(i\partial_2\psi_L^{(1)})P_L^{(1)}+\psi_L^{(1)*}P_L^{(1)*}(i\partial_2\psi_L^{(2)})P_L^{(2)}] \nonumber\\
&&+\int d^{5}x \frac{\sqrt{g}}{a^{2}(r)} [\psi_R^{(1)*}P_R^{(1)*}(i\partial_3\psi_R^{(1)})P_R^{(1)}-\psi_R^{(2)*}P_R^{(2)*}(i\partial_3\psi_R^{(2)})P_R^{(2)}\nonumber\\
&&-\psi_L^{(1)*}P_L^{(1)*}(i\partial_3\psi_L^{(1)})P_L^{(1)}+\psi_L^{(2)*}P_L^{(2)*}(i\partial_3\psi_L^{(2)})P_L^{(2)}] \nonumber\\
&&-2k\int d^{5}x \frac{\sqrt{g}}{a^{2}(r)} [\psi_L^{(1)*}P_L^{(1)*}\psi_R^{(1)}P_R^{(1)}+\psi_L^{(2)*}P_L^{(2)*}\psi_R^{(2)}P_R^{(2)}\nonumber \\
&&-\psi_R^{(1)*}P_R^{(1)*}\psi_L^{(1)}P_L^{(1)}-\psi_R^{(2)*}P_R^{(2)*}\psi_L^{(2)}P_L^{(2)}]\nonumber\\
&&+\int d^{5}x \frac{\sqrt{g}}{a^{2}(r)} [\psi_L^{(1)*}P_L^{(1)*}\psi_R^{(1)}\partial_rP_R^{(1)}+\psi_L^{(2)*}P_L^{(2)*}\psi_R^{(2)}\partial_rP_R^{(2)}\nonumber \\
&&-\psi_R^{(1)*}P_R^{(1)*}\psi_L^{(1)}\partial_rP_L^{(1)}-\psi_R^{(2)*}P_R^{(2)*}\psi_L^{(2)}\partial_rP_L^{(2)}] .
\end{eqnarray}
We may choose the set of the localization conditions as the following ($i=1,2$)
\begin{subequations}
	\begin{eqnarray}
	\int_0^{\infty} dr a^{3}(r)P_L^{(i)*}P_L^{(i)}=\int_0^{\infty} dr a^{3}(r)P_R^{(i)*}P_R^{(i)}=1 ; \label{eq9a}\\
	\int_0^{\infty} dr a^{3}(r)P_L^{(1)*}P_L^{(2)}=\int_0^{\infty} dr a^{3}(r)P_L^{(2)*}P_L^{(1)}\nonumber\\
	=\int_0^{\infty} dr a^{3}(r)P_R^{(1)*}P_R^{(2)}=\int_0^{\infty} dr a^{3}(r)P_R^{(2)*}P_R^{(1)}=1; 
	\label{eq9b}\\
	-2k	\int_0^{\infty} dr a^{3}(r)P_{L}^{(i)*}(r)P_{R}^{(i)}(r)+	\int_0^{\infty} dr a^{3}(r)P_{L}^{(i)*}(r)\partial_r P_R^{(i)}(r)=-M; \label{eq9c}\\
	2k\int_0^{\infty} dr a^{3}(r)P_{R}^{(i)*}(r)P_{L}^{(i)}(r)-	\int_0^{\infty} dr a^{3}(r)P_{R}^{(i)*}(r)\partial_r P_L^{(i)}(r)=-M. \label{eq9d}
	\end{eqnarray}
\end{subequations}
Note that \eqref{eq9a} and \eqref{eq9b} give $P_{L}^{(1)}=P_{L}^{(2)}$ and $P_{R}^{(1)}=P_{R}^{(2)}$.

The Euler-Lagrange equation corresponding to the eq. \eqref{eq6}, the covariant derivative in Ref. \cite{Jones}, the  Dirac equation $i\gamma^\mu\partial_\mu\psi(x)=m\psi(x)$ in the 4D Minkowski spacetime give the equation for  $P_{R,L}$:
\begin{subequations}
	\begin{eqnarray}
	mP_R^{(i)}-2kP_R^{(i)}+\partial_rP_R^{(i)}=0, \label{eq10a}\\
	mP_L^{(i)}+2kP_L^{(i)}-\partial_rP_L^{(i)}=0. \label{eq10b}
	\end{eqnarray}
\end{subequations}
with their solutions
\begin{subequations}
	\begin{eqnarray}
	P_R^{(1)}=P_R^{(2)}=b_{1/2}e^{r(-m+2k)}, \label{eq11a}\\
	P_L^{(1)}=P_L^{(2)}=d_{1/2}e^{r(m+2k)}. \label{eq11b}
	\end{eqnarray}
\end{subequations}
where $b_{1/2}$ and $d_{1/2}$ are integration constants. Inserting these solutions to the localization conditions  \eqref{eq9a}-\eqref{eq9b} for right and left handed spinors respectively  we get
\begin{subequations}
	\begin{eqnarray}
	|{b_{1/2}}|^{2}=2m-k,\text{ $k<2m$},\label{eq12a} \\
	|{d_{1/2}}|^{2}=2m+k, \text{ $k<-2m$}, \label{eq12b}
	\end{eqnarray}
\end{subequations}
while equations \eqref{eq11a}-\eqref{eq11b} require $k<0$ for finite values of the constants $M$. Intersecting the above conditions gives $k<0$ where $m$ can be zero or positive. Thus one concludes that both the right- and left-handed are localizable for increasing warp factor $k<0$.

\section{LOCALIZATION OF INTERACTING FIELDS}

References \cite{Jones, Triyanta} reported the investigation of localization properties of fundamental matter fields of various spins (spin-0, spin-1/2, and spin-1) that do not interact with other fields except with gravity. Their interactions with gravity are introduced through the background RS and MRS metrics. In reality, most matter fields interact with other fields, in addition to interaction with gravity, where the interaction is defined through a gauge mechanism. Examples such interacting fields are  scalar field-photon coupling, spinor field-photon coupling  and interaction among spin-1 gluon (vector) fields, as described by the Yang-Mills theory. Accordingly, it is necessary to expand the investigation in Refs. \cite{Jones, Triyanta} on field localization by considering the following systems of interacting fields: scalar-vector, spinor-vector, and vector-vector fields.  We derive general localization conditions for each system and check whether the conditions are satisfied by  solutions of field equations corresponding to extra dimension. Here we get the result that each considered system is not localizable in the RS model. On the contrary all systems are localizable in the MRS model.

\subsection{Scalar-Vector Fields}

We first investigate the localization properties of a system of  massless scalar and vector fields coupled through a gauge mechanism. The system is described by the action:
\begin{eqnarray}\label{eq13}
S &=& \int d^{5}x\sqrt{g}\biggl[(\partial_M-ieA_M)\Phi^{*}(\partial^{M}+ieA^{M})\Phi-\frac{1}{4}F_{MR}F^{MR}\biggr],
\end{eqnarray}
where $\emph{g}$ is the determinant of the metric on which is dealing with i.e. the RS or MRS model, $F_{MR}=\partial_MA_R-\partial_RA_M$ is the five-dimensional Faraday tensor with the five-dimensional gauge vector field $A_M$. 

The corresponding field equations are
\begin{eqnarray}\label{eq14}
\partial_M(\sqrt{g}g^{MN}\partial_N\Phi)+ie\partial_M(\sqrt{g}g^{MN}A_N\Phi)+ie\sqrt{g}g^{MN}\partial_N\Phi A_M\nonumber \\-e^2\sqrt{g}g^{MN}A_MA_N\Phi=0,
\end{eqnarray}
\begin{eqnarray}\label{eq15}
\partial_{P}(\sqrt{g}F^{PM})+ie\sqrt{g}\Phi\partial^{M}\Phi^{*}-ie\sqrt{g}\Phi^{*}\partial^{M}\Phi-2e^2\sqrt{g}A^{M}\Phi^{*}\Phi=0.
\end{eqnarray}

Decomposing $A_M =(A_\mu(x^{M}), A_5)=(a_\mu(x^{\mu})c(y), A_y)$ and $\Phi(x^{M})=\varphi (x^{\mu})\chi(x^{5})$ and taking $A_5=A_y=const$ as in Ref. \cite{Jones}, the action becomes
\begin{eqnarray}\label{eq16}
S &=& \int_{0}^{\infty} dx^5\sqrt{g}\chi^{*}\chi g^{\mu\nu}\int d^{4}x \partial_\mu \varphi ^{*} \partial_\nu\varphi+\int_{0}^{\infty} dx^5\sqrt{g}g^{55}\partial_5\chi^{*}\partial_5\chi \int d^{4}x \varphi ^{*}\varphi\nonumber\\
&&+\int_{0}^{\infty} dx^5 \sqrt{g} c(x^5)\chi^{*}\chi g^{\mu\nu}\int d^{4}x ie\partial_\mu \varphi ^{*}a_\nu\varphi \nonumber\\
&&+\int_{0}^{\infty} dx^5\sqrt{g}g^{55}\partial_5\chi^{*}A_5\chi\int d^{4}x ie\varphi ^{*}\varphi \nonumber\\
&&+\int_{0}^{\infty} dx^5 \sqrt{g} c(x^5)\chi^{*}\chi g^{\mu\nu}\int d^{4}x (-ie)\partial_\nu\varphi a_\mu   \varphi^{*}\nonumber \\
&&+\int_{0}^{\infty} dx^5\sqrt{g}g^{55}\chi^{*}\partial_5\chi A_5 \int d^{4}x (-ie)\varphi\varphi^{*}\nonumber\\
&&+\int_{0}^{\infty} dx^5\sqrt{g}c^{2}(x^5)\chi^{*}\chi g^{\mu\nu}\int d^{4}x e^{2}a_\mu a_\nu\varphi ^{*}\varphi\nonumber \\
&&+ \int_{0}^{\infty} dx^5\sqrt{g}g^{55}\chi^{*}\chi A_5A_5\int d^{4}x e^{2}\varphi ^{*}\varphi \nonumber\\
&&+\int_{0}^{\infty} dx^5 \sqrt{g}c^{2}(x^5)g^{\mu\nu}g^{\alpha\beta}(-\frac{1}{4})\int d^{4}x f_{\mu\alpha}f_{\nu\beta}\nonumber \\
&&+2\int_{0}^{\infty} dx^5\sqrt{g} (\partial_5c(x^5))^{2}g^{\mu\nu}g^{55}(-\frac{1}{4})\int d^{4}x a_{\mu}(x^\nu)a_{\nu}(x^\nu),
\end{eqnarray}
giving the localization conditions for the fields:
\begin{subequations}
\begin{eqnarray}
\begin{split}
\int_0^\infty dx^5\sqrt{g}\chi^{*}\chi g^{\mu\nu}=N_1\eta^{\mu\nu} ; \\
\int_0^\infty dx^5 c(x^5)\sqrt{g}\chi^{*}\chi g^{\mu\nu}=N_2\eta^{\mu\nu};
\int_0^\infty dx^5c^{2}(x^5)\sqrt{g}\chi^{*}\chi g^{\mu\nu}=N_3\eta^{\mu\nu},\label{eq17a}
\end{split}
\end{eqnarray}
\begin{eqnarray}
\int_0^\infty  dx^5 \sqrt{g}g^{55}[(\partial_5\chi^{*})(\partial_5\chi)+ie(\partial_5\chi^{*})A_5\chi]\nonumber \\
+\int_0^\infty  dx^5 \sqrt{g}g^{55}[(-ie)\chi^{*}\partial_5\chi A_5+e^{2}\chi^{*}\chi {A_5}^{2}]
=-{m_\varphi}^{2}, \label{eq17b}\\
\int_0^\infty dx^5 \sqrt{g} c^{2}g^{\mu\nu}g^{\alpha\beta}=N_4\eta^{\mu\nu}\eta^{\alpha\beta}; \int_0^\infty dx^5 \sqrt{g}(\partial_5 c)^{2}g^{\mu\nu}g^{55}=\eta^{\mu\nu}{m_A}^{2}. \label{eq17c}
	\end{eqnarray}
\end{subequations}
In the above, $\varphi$ and $a_\mu$ represent scalar and vector fields on the brane with their masses are $m_\varphi$ and $m_A$ respectively.

The field equations on the other hand, can be written as
\begin{eqnarray}\label{eq18}
\frac{1}{\chi}\partial_{5}(\sqrt{g}g^{55}\partial_5\chi)+ie\frac{1}{\chi}\partial_5(\sqrt{g}g^{55}A_5\chi)+ie\frac{1}{\chi}\sqrt{g}g^{55}A_5\partial_5\chi-e^2\sqrt{g}g^{55}A_5A_5\nonumber \\
=\sqrt{g}g^{\mu\nu}(-\frac{1}{\varphi}\partial_\mu\partial_\nu\varphi-iec\frac{1}{\varphi}\partial_\mu(a_\nu\varphi)-iec\frac{1}{\varphi}a_\mu\partial_\nu\varphi+e^2c^2a_\mu a_\nu),
\end{eqnarray}
\begin{eqnarray}\label{eq19}
0 &=& c(x^5)\partial_{\mu} (\sqrt{g}g^{\mu\nu}g^{M\alpha}f_{\nu\alpha})-\partial_{\mu}(\sqrt{g}g^{\mu\nu}g^{M5}a_{\nu}(x^\mu)\partial_5c(x^5)\nonumber \\
&&+\partial_5(\sqrt{g}g^{55}g^{M\alpha}a_{\alpha}(x^\mu)\partial_5c(x^5))\nonumber \\
&& +\sqrt{g}g^{M\alpha}\chi^{*}\chi[ie\varphi\partial_\alpha\varphi^*-ie\varphi^*\partial_\alpha\varphi+2e^2c(x^5)a_\alpha(x^\mu)\varphi^*\varphi]\nonumber\\
&& +\sqrt{g}g^{M5}[ie\varphi^{*}\varphi\chi\partial_5\chi^*-ie\varphi^*\varphi\chi^*\partial_5\chi+2e^2A_5\chi^*\chi\varphi^*\varphi].
\end{eqnarray}

The fulfillment of the fields on the localization conditions depends on the functions $c(x^5)$ and $\chi(x^5)$, the extra-dimension part solution of the field equations. The function $c(x^5)$ can be deduced from the first field equation \eqref{eq18}. The LHS of this equation depends only on the extra coordinate $x^5$  while the RHS depends on all coordinates where all non-extra coordinates are fully inside the bracket. The dependence of all terms in the bracket on the extra coordinate is given by the functions  $c(x^5)$ and $c^2(x^5)$. Thus considering all coordinates are independent one another the function $c(x^5)$ should be a constant. Another possibility is that the scalar $\varphi$ and vector $a_\nu$  field functions should be such that the non-extra coordinates disappear in the equation. For example, $\varphi$ is an exponential function and $a_\nu$ is a constant. However this is just a special case for $\varphi$ and $a_\nu$ and we do not take this choice into account. For $c(x^5)$=const  and take the constant equals unity the field equation for the vector field simplifies into
\begin{eqnarray}\label{eq20}
0 &=&  \sqrt{g}g^{\mu\nu}g^{M\alpha}\partial_{\mu}f_{\nu\alpha}
+\sqrt{g}g^{M\alpha}\chi^{*}\chi[ie\varphi\partial_\alpha\varphi^*-ie\varphi^*\partial_\alpha\varphi+2e^2a_\alpha(x^\mu)\varphi^*\varphi]\nonumber\\
&& +\sqrt{g}g^{M5}[ie\varphi^{*}\varphi\chi\partial_5\chi^*-ie\varphi^*\varphi\chi^*\partial_5\chi+2e^2A_5\chi^*\chi\varphi^*\varphi].
\end{eqnarray}
For $M=\beta=0,1,2,3$ and $M=5$ respectively we have
\begin{subequations}
	\begin{eqnarray}
	\partial_{\mu}f^{\mu\beta}
	+a^{2}(y)\chi^{*}\chi[ie\varphi\partial^\beta\varphi^*-ie\varphi^*\partial^\beta\varphi+2e^2a^\beta\varphi^*\varphi]=0,\label{eq21a}
	\end{eqnarray}
	\begin{eqnarray}
	ie(\partial_5\chi^*)\chi-ie\chi^*(\partial_5\chi)+2e^2A_5\chi^*\chi=0. \label{eq21b}
	\end{eqnarray}
\end{subequations}

Until now, all we have discussed is general. It applies to any models. It is time now to look at the RS model characterized by $\sqrt{g}=a^4(y)$, $g^{\mu\nu}=a^{-2}(y)\eta^{\mu\nu}$, $g^{55}=-1$. The localization condition related to the constant  $N_4$  then becomes 
\begin{eqnarray}
\int_{0}^{\infty}dyc^2(y)=N_4. \label{eq22}
\end{eqnarray}
As $c(x^5)$=const  the above integral gives $N_4=\infty$. This means that scalar fields interacting with vector fields are not localizable in the RS model. 

For the MRS model where  $\sqrt{g}=a^5(y)$, $g^{\mu\nu}=a^{-2}(y)\eta^{\mu\nu}$, $g^{55}=-a^{-2}(y)$. The localization conditions, after recalling that $c(x^5)$=const$\equiv$1 reduce into
\begin{subequations}
	\begin{eqnarray} \label{eq23a}
	\int_0^\infty dra^{3}(r)\chi^{*}(r)\chi(r) =N_1=N_2=N_3 \text{,} \int_0^\infty dra(r)=N_4  \text{, $m_A=0$},
	\end{eqnarray}
	\begin{eqnarray} \label{eq23b}
	-\int_0^\infty dr a^3(r)[\partial_r\chi^{*}\partial_r\chi+ie\partial_r\chi^{*}A_r\chi+(-ie)\chi^{*}\partial_r\chi A_r+e^{2}\chi^{*}\chi A_r^{2}]=-m_\varphi^{2}.
	\end{eqnarray}
\end{subequations}
while the field equations become
\begin{eqnarray}\label{eq24}
\frac{1}{\chi}\partial_{5}\partial_{5}\chi-3k\frac{1}{\chi}\partial_{5}\chi+2ieA_{5}\frac{1}{\chi}\partial_{5}\chi-3kieA_{5}-e^{2}A_5A_5\nonumber\\
=\eta^{\mu\nu}(\frac{1}{\varphi}\partial_\mu\partial_{\nu}\varphi+ie\frac{1}{\varphi}\partial_\mu(a_\nu\varphi)+ie\frac{1}{\varphi}a_\mu\partial_\nu\varphi-e^{2}a_\mu a_\nu).
\end{eqnarray}
\begin{subequations}
	\begin{eqnarray}\label{eq25a}
	0=\partial_{\mu}f^{\mu\beta}+a^{2}(r)\chi^{*}\chi \biggl[ie\varphi\partial^{\beta}\varphi^{*}-ie\varphi^{*}\partial^{\beta}\varphi+2e^{2}a^{\beta}(x^{\mu})\varphi^{*}\varphi\biggr].
	\end{eqnarray}
	\begin{eqnarray} \label{eq25b}
	0=ie\chi\partial_5\chi^*-ie\chi^*\partial_5\chi+2e^2A_5\chi^*\chi.
	\end{eqnarray}
\end{subequations}
The second equation of motion \eqref {eq25a} gives $a^{2}(r)\chi^{*}\chi$=const giving $\chi^*\chi$=const $e^{2kr}$  guarantying the finiteness of $N_1$, $N_2$, $N_3$ and $N_4$. All these constants depend on the parameter $k$: $N_1=N_2=N_3$=const/$k$, $N_4=1/k$. In the first field equation \eqref{eq24}, the LHS depends only on the extra coordinate while the RHS does not depend on the extra coordinate. This means that the LHS is a constant. Since the RHS reminds us to the Klein-Gordon equation of the scalar field interacting with a gauge field in a four-dimensional Minkowski space the constant is nothing but the quadratic mass of the scalar field ${m_\varphi}^2$. Thus
\begin{eqnarray}\label{eq26}
\frac{1}{\chi}\partial_{5}\partial_{5}\chi-3k\frac{1}{\chi}\partial_{5}\chi+2ieA_{5}\frac{1}{\chi}\partial_{5}\chi-3kieA_{5}-e^{2}A_5A_5=-{m_\varphi}^2.
\end{eqnarray}
The general solution is ($b_0$ and $c_0$ are constants of integration)
\begin{eqnarray}\label{eq27}
\chi(r)=b_0 exp \biggl\{ [\frac{3}{2}k-ieA_{5}+\sqrt {(9/4)k^2-m_{\varphi}^2}]r\biggr\}\nonumber \\+c_0 exp \biggl\{ [\frac{3}{2}k-ieA_{5}-\sqrt {(9/4)k^2-m_{\varphi}^2}]r\biggr\}.
\end{eqnarray}
This solution matches with the previous result, i.e. $\chi^*\chi=$const $e^{2kr}$,  for the case of $b_0=0$  and $m_\varphi=\sqrt{2}k$:
\begin{eqnarray}\label{eq28}
\chi(r)=c_{0}exp \{r(k-ieA_5)\}.
\end{eqnarray}
This result is also in accordance with the last field equation \eqref{eq26}. Thus unlike in the RS model, scalar fields interacting with vector fields are localizable in the MRS model for decreasing warp factor.

\subsection{Non-Abelian Field}

Next, we investigate the localization  of  a system of vector fields coupled to themselves in  five-dimensional RS and MRS models. The system is described by a Lagrangian introduced in Yang-Mills theory:
\begin{equation}
\label{eq29}
\mathcal{L}=-\frac{1}{4}\sqrt{g}W_{AB}^{a}  W ^{aAB},
\end{equation}
where $ W_{AB}^{a}=\partial_{A}W_{B}^{a}-\partial_{B}W_{A}^{a}+hf^{abc}W_{A}^{b}W_B^{c}$ are field strengths, $h$ is a coupling constant and $f^{abc}$ is a structure constant. $W_{A}^{a} $  is a vector in an internal symmetry space with $n^{2}-1$ dimension and its components are defined by $W_{A}^{1},...,W_{A}^{n^{2}-1}$ \cite{Ryder}. The third term of the five-dimensional tensor shows that a vector field interacts with another vector field. We decompose $W_{A}^{a}=(w_{\alpha}^{a}(x^\beta)c(x^5),W_5^{a})$ and  choose $W_5^{a}$  a constant as in previous reference \cite{Jones}, so that  the five-dimensional action corresponding to the Lagrangian \eqref{eq29} can be written as the following:
\begin{eqnarray}
S &=&-\frac{1}{4}\int_{0}^{\infty} dx^5\sqrt{g} c^{2}(x^5)g^{\alpha\nu}g^{\beta\sigma}\int d^{4}x(\partial_{\alpha}w_{\beta}^{a}-\partial_{\beta}w_{\alpha}^{a})(\partial_{\nu}w_{\sigma}^{a}-\partial_{\sigma}w_{\nu}^{a})\nonumber\\
&&-\frac{1}{2}\int_{0}^{\infty} dx^5 \sqrt{g} c^{3}(x^5)g^{\alpha\nu}g^{\beta\sigma}\int d^{4}xh(\partial_{\alpha}w_{\beta}^{a}-\partial_{\beta}w_{\alpha}^{a})f^{ade}w_{\nu}^{d}w_{\sigma}^{e}\nonumber\\
&&-\frac{1}{4}\int_{0}^{\infty} dx^5 \sqrt{g} c^{4}(x^5)g^{\alpha\nu}g^{\beta\sigma}\int d^{4}xh^{2}f^{abc}w_{\alpha}^{b}w_{\beta}^{c}f^{ade}w_{\nu}^{d}w_{\sigma}^{e}\nonumber\\
&&-\frac{1}{2}\int_{0}^{\infty} dx^5 \sqrt{g} (\partial_{5}c(x^5))^{2}g^{\alpha\nu}g^{55}\int d^{4}x w_{\alpha}^{a}w_{\nu}^{a}\nonumber\\
&&+\int_{0}^{\infty}dx^5\sqrt{g}c(x^5)\partial_{5}c(x^5)g^{\alpha\nu}g^{55}W_{5}^{c}\int d^{4}xhf^{abc}w_{\alpha}^{b}w_{\nu}^{a} \nonumber \\
&&-\frac{1}{2}\int_{0}^{\infty} dx^5 \sqrt{g}c^{2}(x^5)g^{\alpha\nu}g^{55}W_{5}^{c}W_{5}^{e}\int d^{4}x h^{2}f^{abc}w_{\alpha}^{b}f^{ade}w_{\nu}^{d}\label{eq30}.
\end{eqnarray}
The localization conditions  are the following
\begin{subequations}
	\begin{eqnarray}
	\int_{0}^{\infty} dx^5\sqrt{g} c^{2}g^{\alpha\nu}g^{\beta\sigma}=N_1\eta^{\alpha\nu}\eta^{\beta\sigma};\int_{0}^{\infty} dx^5 \sqrt{g} c^{3}g^{\alpha\nu}g^{\beta\sigma}=N_2\eta^{\alpha\nu}\eta^{\beta\sigma};\nonumber\\
	\int_{0}^{\infty} dx^5 \sqrt{g} c^{4}g^{\alpha\nu}g^{\beta\sigma}=N_3\eta^{\alpha\nu} \eta^{\beta\sigma},\label{eq31a}\\
	-\frac{1}{2}\int_{0}^{\infty} dx^5 \sqrt{g} (\partial_{5}c)^{2}g^{\alpha\nu}g^{55}=-\frac{1}{2}{m_{w}}^{2}\eta^{\alpha\nu},\label{eq31b}\\
	\int_{0}^{\infty}dx^5\sqrt{g}c(\partial_{5}c)g^{\alpha\nu}g^{55}W_{5}^{c}=N_4\eta^{\alpha\nu}; \nonumber \\ -\frac{1}{2}\int_{0}^{\infty} dx^5 \sqrt{g}c^{2}g^{\alpha\nu}g^{55}W_{5}^{c}W_{5}^{e}=N_5\eta^{\alpha\nu}\label{eq31c}.
	\end{eqnarray}
\end{subequations}
where the constants $N_1, N_2, N_3, N_4, N_5$ and $m_w$ are finites.  

The equation of motion for the non-abelian vector field corresponding to the Lagrangian \eqref{eq29} reads
\begin{eqnarray}
&&-\sqrt{g}g^{\alpha\sigma}g^{\beta Q}\partial_\sigma W_{\alpha\beta}^{i}-\sqrt{g}g^{\alpha\sigma}g^{5Q}\partial_\sigma W_{\alpha 5}^{i}+\partial_5[-\sqrt{g}g^{55}g^{\beta Q}W_{5\beta}^{i}] \nonumber \\
&&+\sqrt{g}g^{Q\alpha}g^{\nu\beta}hf^{aie}W_{\nu}^{e}W_{\alpha\beta}^{a}+\sqrt{g}g^{Q\alpha}g^{55}hf^{aie}W_{5}^{e}W_{\alpha 5}^{a}\nonumber\\ &&+\sqrt{g}g^{Q5}g^{\nu\beta}hf^{aie}W_{\nu}^{e}W_{5\beta}^{a}=0\label{eq32}.
\end{eqnarray}

In the RS model, $Q=5$ gives $\partial^{\alpha}W_{\alpha 5}^{i}=0$ while  $Q=\lambda=0,1,2,3$ give
\begin{eqnarray}
-\eta^{\alpha\sigma}\eta^{\beta\lambda}\partial_\sigma[(\partial_\alpha w_{\beta}^{i}-\partial_\beta w_{\alpha}^{i})c(y)+hf^{ide}c^2(y)w_{\alpha}^{d}w_{\beta}^{e}] \nonumber \\
-2ka^2(y)\eta^{\beta\lambda}[\partial_yc(y)w_{\beta}^{i}+hf^{ide}W_{5}^{d}c(y)w_{\beta}^{e}]\nonumber \\
+a^{2}(y)\eta^{\beta\lambda}[\partial_{y}^2c(y)w_{\beta}^{i}+hf^{ide}W_{5}^{d}\partial_yc(y)w_{\beta}^{e}]\nonumber \\
+\eta^{\lambda\alpha}\eta^{\nu\beta}hf^{aie}c(y)w_{\nu}^{e}[(\partial_\alpha w_{\beta}^{a}-\partial_\beta w_{\alpha}^{a})c(y)+hf^{abc}c^{2}(y)w_{\alpha}^{b}w_{\beta}^{c}]\nonumber\\
-a^2(y)\eta^{\lambda\alpha}hf^{aie}W_{5}^{e}[-\partial_yc(y)w_{\alpha}^{a}+hf^{abc}c(y)w_{\alpha}^{b}W_{5}^{c}]=0. \label{eq33}
\end{eqnarray}
Each term of the above equation contains different functions of extra coordinate $x^5=y$, namely $c(y)$, $c^2(y)$, $c^3(y)$, $a^2(y)c(y)$, $a^2(y)\partial_yc(y)$, and $a^2(y)\partial_y^2c(y)$. Since $y$ is independent with $x^\mu$ all those functions should be constant. This can only be fulfilled with $c=0$ or $c=$constant with $k=0$. The former means that there is no YM field $W_{\mu}^{a}$ (but still we may have $w_{\mu}^{a}$) while the later means that the spacetime is Minkowskian. Thus we are not able to discuss localization of YM fields in the RS model.

We now discuss the Yang Mills field in the MRS model. Recalling for the MRS model that $\sqrt{g}=a^5(r)$, $g^{\mu\nu}=a^{-2}(r)\eta^{\mu\nu}$, $g^{55}=-a^{-2}(r)$  the localization conditions give integrands in \eqref{eq31a}-\eqref{eq31c} differ in the power of $a(x^5)$ compared to the RS model. This leads to different conclusion on localization as will be shown below. To see whether all  $N_i$ in \eqref{eq31a}-\eqref{eq31c} are finite we should derive the function $c(r)$ from the field equations \eqref{eq32}. In the MRS model, eq. \eqref{eq32} for $Q=5$ give $\partial^{\alpha}W_{\alpha 5}^{i}=0$ as in the RS model while for $Q=\lambda=0, 1, 2, 3$, give
\begin{eqnarray}
-\eta^{\alpha\sigma}\eta^{\beta\lambda}\partial_\sigma[(\partial_\alpha w_{\beta}^{i}-\partial_\beta w_{\alpha}^{i})c(r)+hf^{ide}c^2(r)w_{\alpha}^{d}w_{\beta}^{e}] \nonumber \\
-k\eta^{\beta\lambda}[\partial_rc(r)w_{\beta}^{i}+hf^{ide}W_{5}^{d}c(r)w_{\beta}^{e}]\nonumber \\
+\eta^{\beta\lambda}[\partial_{r}^2c(r)w_{\beta}^{i}+hf^{ide}W_{5}^{d}\partial_rc(r)w_{\beta}^{e}]\nonumber \\
+\eta^{\lambda\alpha}\eta^{\nu\beta}hf^{aie}c(r)w_{\nu}^{e}[(\partial_\alpha w_{\beta}^{a}-\partial_\beta w_{\alpha}^{a})c(r)+hf^{abc}c^{2}(r)w_{\alpha}^{b}w_{\beta}^{c}]\nonumber\\
-\eta^{\lambda\alpha}hf^{aie}W_{5}^{e}[-\partial_rc(r)w_{\alpha}^{a}+hf^{abc}c(r)w_{\alpha}^{b}W_{5}^{c}]=0. \label{eq34}
\end{eqnarray}

Unlike in \eqref{eq33} there is no function $a(r)$ in \eqref{eq34}. Here the function of the extra coordinate $r$ found in \eqref{eq34} are of the forms $c(r)$, $c^2(r)$, $c^3(r)$, $\partial_{r}c(r)$, and $\partial_{r}^{2}c(r)$. Thus $c(r)$ should be a constant since $r$ and $x^\mu$ are independent one another. Inserting $c(r)=$constant into \eqref{eq31a}-\eqref{eq31c} gives all $N_{i}$ are finite and $m_w=0$ for $k>0$. Thus in conclusion, the non-abelian vector field is localizable on the brane $r=0$ for the case of massless mode and a decreasing warp factor.

\subsection{Spinor-Vector Fields}

Next, we investigate the localization  of  a system of  massless spinor fields and vector fields coupled through a gauge mechanism. The system is described by the action:
\begin{eqnarray}
S=\int d^{5}x \sqrt{g}[\overline{\Psi}i\Gamma^MD_M\Psi-\frac{1}{4}F_{MR}F^{MR}], \label{eq35}
\end{eqnarray}
where $D_M$ is the covariant derivative, $D_M=\partial_M-ieA_M+\frac{1}{4}\omega_{M}^{\overline{M}\overline{N}} \sigma_{\overline{M}\overline{N}}$  with  $\sigma_{\overline{M}\overline{N}}$ and  $\omega_{M}^{\overline{M}\overline{N}}$ defined in Ref. \cite{Jones}. The corresponding covariant derivatives for general metric take form 
\begin{subequations}
\begin{eqnarray}
D_\mu&=&\partial_\mu-ieA_\mu+\frac{ik}{2} \frac{a}{b}\gamma_{\overline{\mu}}\gamma_{\overline{5}}, \label{eq36a}\\ 
&&D_5=\partial_5-ieA_5, \label{eq36b}
\end{eqnarray}
\end{subequations}
where  $b=1$ in the RS model and  $b=a=exp(-kx^5)$ in the MRS model. 

Decomposing the five-dimensional spinor $\Psi(x^{M})$ as in \eqref{eq7} the expression of the action becomes so lengthy and it shows variety of integrals over the extra coordinate explicitly leading the following localization conditions ($i$=1,2):
\begin{subequations}
	\begin{eqnarray}
	\int_0^{\infty} dx^5 {\sqrt{g}}e_{\overline 0}^{0}e_{\overline 0}^{0}P_L^{(i)*}P_L^{(i)}=N_{L0}^{(i)}, \text{L}\leftrightarrow \text{R}; \label{eq37a}\\
	\int_0^{\infty} dx^5 {\sqrt{g}}e_{\overline 0}^{0}e_{\overline 0}^{0}cP_L^{(i)*}P_L^{(i)}=M_{L0}^{(i)}, \text{L}\leftrightarrow \text{R};\label{eq37b}\\
	\int_0^{\infty} dx^5 {\sqrt{g}}e_{\overline 0}^{0}e_{\overline 1}^{1}P_L^{(i)*}P_L^{(j)}=N_{L1}^{(i,j)}, \text{L}\leftrightarrow \text{R}, \text{i}\neq\text{j};\label{eq37c}  \\ 
	\int_0^{\infty} dx^5 {\sqrt{g}}e_{\overline 0}^{0}e_{\overline 1}^{1}cP_L^{(i)*}P_L^{(j)}=M_{L1}^{(i,j)}, \text{L}\leftrightarrow \text{R}, \text{i}\neq\text{j};\label{eq37d}\\
	\int_0^{\infty} dx^5 {\sqrt{g}}e_{\overline 0}^{0}e_{\overline 2}^{2}P_L^{(i)*}P_L^{(j)}=N_{L2}^{(i,j)}, \text{L}\leftrightarrow \text{R}, \text{i}\neq\text{j}; \label{eq37e}  \\
	\int_0^{\infty} dx^5 {\sqrt{g}}e_{\overline 0}^{0}e_{\overline 2}^{2}cP_L^{(i)*}P_L^{(j)}=M_{L2}^{(i,j)}, \text{L}\leftrightarrow \text{R}, \text{i}\neq\text{j}; \label{eq37f}\\
	\int_0^{\infty} dx^5 {\sqrt{g}}e_{\overline 0}^{0}e_{\overline 3}^{3}P_L^{(i)*}P_L^{(i)}=N_{L3}^{(i)}, \text{L}\leftrightarrow \text{R}; \label{eq37g}\\
	\int_0^{\infty} dx^5 {\sqrt{g}}e_{\overline 0}^{0}e_{\overline 3}^{3}cP_L^{(i)*}P_L^{(i)}=M_{L3}^{(i)}, \text{L}\leftrightarrow \text{R};  \label{eq37h}\\
	-2k\int_0^{\infty} dx^5 {\sqrt{g}}e_{\overline 0}^{0}\frac{1}{b}P_L^{(i)*}P_R^{(i)}=M_{LRb}^{(i)}, \text{L}\leftrightarrow \text{R}; 
	\label{eq37i}\\
	\int_0^{\infty} dx^5 {\sqrt{g}}e_{\overline 0}^{0}e_{\overline 5}^{5}P_L^{(i)*}\partial_5P_R^{(i)}=N_{LR}^{(i)}, \text{L}\leftrightarrow \text{R};  \label{eq37j}\\
	ieA_5\int_0^{\infty} dx^5 {\sqrt{g}}e_{\overline 0}^{0}e_{\overline 5}^{5}P_{L}^{(i)*}P_{R}^{(i)}=M_{LR}^{(i)}, \text{L}\leftrightarrow \text{R}; \label{eq37k}\\
	\int_0^{\infty} dx^5 \sqrt{g} g^{\alpha\nu} g^{\beta\sigma}c^{2}=N_{1}\eta^{\alpha\nu}\eta^{\beta\sigma};
	\int_0^{\infty} dx^5 \sqrt{g}g^{\alpha\nu}g^{55}(\partial_5 c)^{2}=-m_A\eta^{\alpha\nu}. \label{eq37l}
	\end{eqnarray}
\end{subequations}
All quantities on the RHS of the above conditions are finite and depend on the warp parameter $k$. Note that the sum of the integrals containing both $P_L$ and $P_R$ altogether, up to some constant factors, correspond to the mass of the spinor field, namely,
\begin{eqnarray}
\int_{0}^{\infty}dx^5\sqrt{g} e_{\overline 0}^{0}e_{\overline 5}^{5}P_L^{(i)*}\partial_5P_R^{(i)}+ieA_5\int_0^{\infty} dx^5 {\sqrt{g}}e_{\overline 0}^{0}e_{\overline 5}^{5}P_{L}^{(i)*}P_{R}^{(i)}\nonumber \\
-2k\int_0^{\infty} dx^5 {\sqrt{g}}e_{\overline 0}^{0}\frac{1}{b}P_L^{(i)*}P_R^{(i)}=-m_\psi,                  \text{L}\leftrightarrow \text{R}.	\label{eq38}
\end{eqnarray}
Since $e_{\overline 0}^{0}$= $e_{\overline 1}^{1}$= $e_{\overline 2}^{2}$= $e_{\overline 3}^{3}$=$\frac{1}{a}$ and  $e_{\overline 5}^{5}$=$\frac{1}{b}$ for both RS and MRS models, the conditions simplify into
\begin{subequations}
	\begin{eqnarray}
	\int_0^{\infty} dx^5 \frac{\sqrt{g}}{a^2}P_L^{(i)*}P_L^{(i)}=N_{L0}^{(i)}=N_{L3}^{(i)}, \text{L}\leftrightarrow \text{R};\label{eq39a}\\
	\int_0^{\infty} dx^5 \frac{\sqrt{g}}{a^2}cP_L^{(i)*}P_L^{(i)}=M_{L0}^{(i)}=M_{L3}^{(i)}, \text{L}\leftrightarrow \text{R};\label{eq39b}\\	
	\int_0^{\infty} dx^5\frac{\sqrt{g}}{a^2}P_L^{(i)*}P_L^{(j)}=N_{L1}^{(i,j)}=N_{L2}^{(i,j)}, \text{L}\leftrightarrow \text{R}, \text{i}\neq\text{j};\label{eq39c}\\	
	\int_0^{\infty} dx^5\frac{\sqrt{g}}{a^2}cP_L^{(i)*}P_L^{(j)}=M_{L1}^{(i,j)}=M_{L2}^{(i,j)}, \text{L}\leftrightarrow \text{R}, \text{i}\neq\text{j};\label{eq39d}\\
	-2k\int_0^{\infty} dx^5\frac{\sqrt{g}}{ab}P_L^{(i)*}P_R^{(i)}=M_{LRb}^{(i)}, \text{L}\leftrightarrow \text{R}; \label{eq39e}\\  ieA_5\int_0^{\infty} dx^5\frac{\sqrt{g}}{ab}P_L^{(i)*}P_R^{(i)}=M_{LR}^{(i)},  \space  \text{L}\leftrightarrow \text{R} ;\label{eq39f}\\
	\int_0^{\infty}dx^5\frac{\sqrt{g}}{ab}P_L^{(i)*}\partial_5P_R^{(i)}=N_{LR}^{(i)}, \text{L}\leftrightarrow \text{R} ;\label{eq39g}\\
	\int_0^{\infty}dx^5\frac{\sqrt{g}}{a^4}c=N_1; \int_0^{\infty}dx^5\frac{\sqrt{g}}{a^2b^2}{(\partial_5c)}^2=m_A. \label{eq39h}
	\end{eqnarray}
\end{subequations}

We could not analyze whether the spinor-vector system fulfill the above conditions until we look at the field equations. This is because $P_{L,R}^{(i)}$ and $c$ are solutions of the field equations that correspond to the extra coordinate. Even though the integrands in conditions \eqref{eq39a} and \eqref{eq39b} differ by a factor $c(x^5)$ one could not specify $c(x^5)$. $c(x^5)$=const is consistent with both conditions but this not the only choice. For example, functions $c(x^5)$ of the form $c(x^5)=1+e(x^5){a^2}/(\sqrt{g}P_L^{(i)*}P_{L}^{(i)})$ with $\sqrt{g}P_L^{(i)*}P_{L}^{(i)}\neq 0$ for all $x^5$ and $\int_{0}^{\infty}dx^5e(x^5)=0$ are consistent with the conditions
\eqref{eq39a} and \eqref{eq39b}. As there are many functions fulfilling $\int_{0}^{\infty}dx^5e(x^5)=0$, we understand that $c(x^5)$ is not unique. We must look at the field equations first before going through the localization conditions again.

The field equations corresponding to spinor field are 
\begin{eqnarray}\label{eq40}
i\Gamma^{M}D_M\Psi=\frac{1}{a}i\gamma^{\overline \mu}D_\mu\Psi+\frac{1}{b}i(-i\gamma^{\overline 5})D_5\Psi,\nonumber \\ 
\frac{1}{a}i\gamma^{\overline \mu}(\partial_\mu-ieca_\mu)\Psi+i\frac{1}{a}\gamma^{\overline \mu}(\frac{ik}{2}\frac{a}{b}\gamma_{\overline\mu}\gamma_{\overline5})+\frac{1}{b}i(-i\gamma^{\overline 5})(\partial_5-ieA_5)\Psi=0.
\end{eqnarray}
In the above equations the dependence on the extra coordinate comes from $a$, $b$, $c$ and $\Psi$. The equations make sense only if all these functions have the form  that all terms in the equations are  functions of the extra coordinate which are proportional to one another. Accordingly the function $c(x^5)$ should be a constant. As $c(x^5)$ is a constant the integrand in the first condition \eqref{eq39h} is proportional to $\frac{\sqrt{g}}{a^4}$ which is equal to unity for the RS model. Accordingly the integral is divergent meaning that the RS model disobeys the first condition \eqref{eq39h}. So for the rest of our discussion we only consider the MRS model.

Taking $c(x^5)=1$ the first term of the above equation corresponds to the mass term of the Dirac equation in the four-dimensional Minkowski space, $i\gamma^{\overline \mu}(\partial_\mu-iea_\mu)\psi=m\psi$. Thus we have, after taking $a=b$
\begin{eqnarray}\label{eq41}
0=m\Psi+\gamma^{\overline 5}(\partial_5-ieA_5-2k)\Psi,
\end{eqnarray}
which is equivalent to 
\begin{subequations}
	\begin{eqnarray}
	mP_{R}^{(i)}+(\partial_5-ieA_5-2k)P_{R}^{(i)}=0, \label{eq42a} \\
	mP_{L}^{(i)}-(\partial_5-ieA_5-2k)P_{L}^{(i)}=0. \label{42b} 	
	\end{eqnarray}
\end{subequations}
The solutions are
\begin{subequations}
	\begin{eqnarray}
	P_{R}^{(1)}(r)=P_{R}^{(2)}(r)=b_{1/2}exp[2k-m+ieA_5]r; \label{eq43a}\\	
	P_{L}^{(1)}(r)=P_{L}^{(2)}(r)=d_{1/2}exp[2k+m+ieA_5]r\label{eq43b}	
	\end{eqnarray}
\end{subequations}
where $b_{1/2}$ and $d_{1/2}$ are integration constants. 

The constancy of  $c(x^5)$ leads to $m_A=0$ and $\int_{0}^{\infty}dr a(r)=N_1$ in \eqref{eq39h} meaning the vector field should be massless and the warp factor should decrease, $k>0$, for  localization.
Inserting the solutions \eqref{eq43a} and \eqref{eq43b} into eq. \eqref{eq39a}-\eqref{eq39d}, the LHS of conditions for both right- and left-handed reduce into
\begin{subequations}
	\begin{eqnarray}
	|b_{1/2}|^{2}\int_{0}^{\infty}dr exp[-(2m-k)r], \label{eq44a}\\
	|d_{1/2}|^{2}\int_{0}^{\infty}dr exp[(2m+k)r].\label{eq44b}
	\end{eqnarray}
\end{subequations}
Since $k$ must be positive for localization, the finiteness of (44) gives $m>k/2$ and $d_{1/2}=0$. The latter is equivalent to $P_L=0$. This does not mean that $\psi_L=0$, i.e we still have a complete pair of spinors in a four-dimensional Minkowski spacetime $(\psi_R, \psi_L)$. Conditions \eqref{eq39e}-\eqref{eq39g} give $k<0$ and $d_{1/2}=0$ for localization and only the latter fits with the previous result. Thus, the possible localization until this step is that the spinor field should have $P_L=0$ and the warp parameter $k$ should be positive. Further analysis corresponds to the vector field equation:
\begin{eqnarray}
\partial_\mu(\sqrt{g}g^{\mu\alpha}g^{Q\beta}F_{\alpha\beta})+\partial_\mu(\sqrt{g}g^{\mu\alpha}g^{Q5}F_{\alpha 5})\nonumber \\
+\partial_5(\sqrt{g}g^{55}g^{Q\beta}F_{5\beta})+\sqrt{g}\overline{\Psi}e\Gamma^{Q}\Psi=0 \label{eq45}. 
\end{eqnarray}
Recalling that $A_5$ and $c(x^5)$  are constant the equation for $Q=5$ reduces to $\overline{\Psi} e\gamma^{\overline 5}\Psi=0$  which, for $P_L^{(i)}=0$,  is equivalent to
\begin{eqnarray}
P_L^{(1)*}P_R^{(1)}\psi_L^{(1)*}\psi_R^{(1)}+P_L^{(2)*}P_R^{(2)}\psi_L^{(2)*}\psi_R^{(2)}\nonumber \\-P_R^{(1)*}P_L^{(1)}\psi_R^{(1)*}\psi_L^{(1)}-P_R^{(2)*}P_L^{(2)}\psi_R^{(2)*}\psi_L^{(2)}=0.\label{eq46}
\end{eqnarray}
All terms within the left hand side of the above equation contain $P_L$. Thus $P_L=0$ is in accordance to eq. \eqref{eq45}.

For $Q=\nu=0,1,2,3$ the equation \eqref{eq45} reduces into
\begin{eqnarray}
\partial_\mu(f^{\mu\nu})+a^{2}(r)\overline{\Psi}e\gamma^{\overline \nu}\Psi=0. \label{eq47}
\end{eqnarray}
Since the first term on the left-hand side of eq. \eqref{eq47} does not depend on the extra coordinate, the second term should have the form of $F(r)G(x^\mu)$ where $F(r)=constant$. It is easily proved that for every $\nu=0,1,2,3$, the function $F(r)$ is equal to $a^2 \lvert P_R\rvert^2=constant$ after recalling that $P_L=0$  we have
\begin{eqnarray}
a^2 \lvert P_R\rvert^2 \propto exp2r(k-m)=constant,\label{eq52}
\end{eqnarray}
and $m=k$, which is in accordance to $m>\frac{k}{2}$ from our previous result. So, in conclusion the system of spinor-vector field is localizable on the brane for the case of massive spinors with $m=k$ and of course for $k$ positive.

\section{Conclusions}
First of all, we pointed out a typo in the second equation of (30) in Ref. \cite{Jones} that the power of the warp factor $a(r)$ should be $3$ instead of $1$ (see \eqref{eq5b}). However, correcting this mistake only changes the value of mass in terms of $k$ and does not change the general conclusion on localization that massive scalar fields are localized on the brane both for a decreasing and an increasing warp factors.

Also, we introduced a replacement on the definition of adjoint of spinor fields in ref. \cite{Jones} from $\overline{\Psi}=\Psi^{+}\gamma^{\overline{0}}$ into $\overline{\Psi}=\Psi^{+}\Gamma^{0}$. The later definition of $\overline{\Psi}$ makes more sense since both $\overline{\Psi}$ and $\Gamma^M$ are defined on the same five-dimensional curved spacetime. With the new definition of $\overline{\Psi}$, massive and massless spinor fields are localizable on the brane in the MRS model for an increasing warp factor.

Secondly, we analyzed localizability of interacting fields in the RS and MRS models. Here, a system of interacting fields is said to be localized on the brane if all the fields within the system are localized. The system that we analyzed, i.e  scalar-vector and spinor-vector systems in the RS model were not localizable since the localization conditions  led the integrals over the fifth coordinate in the first equation of  \eqref{eq17c} for scalar-vector system, in the first equation of  \eqref{eq39h} for spinor-vector system were of the form $\int_{0}^{\infty}dy$ which is infinite. While for the vector-vector system the localizability leads to $c(r)=0$ or $c=$constant with $k=0$. This means that one is not able to define localized YM fields in the RS model.

We looked at interacting fields in the MRS model. A system of scalar-vector is localizable  on the brane since all integrals over the extra dimension from $0$ to $\infty$ are finite. We also demonstrated that the condition $a^{2}\chi^{*}\chi$=const from the equation of motion for vector field \eqref{eq25a} gives mass $m_\varphi=\sqrt{2}k$. For a vector-vector system represented by the Yang-Mills field, we obtained that the system is localizable on the brane $r=0$ for the case massless mode and a decreasing warp factor. Finally for the spinor-vector system, we obtained that the system is also localizable on the brane $r=0$ even though with some restrictive conditions: the vector field should be massless, the mass of the spinor field should be $m=k$, $P_L=0$, and the warp factor should decrease. Note that $P_L=0$ does not mean that the field $\psi_L(x^\mu)$ on the brane does not exist.

So, the general conclusion is that in terms of interacting fields, the MRS brane model has better localization properties than the original RS model: the scalar-vector, vector-vector, and spinor-vector systems cannot be localized on the brane $y=0$ in the original RS model while these systems are localizable on the brane $r=0$ (with some restrictions) in the MRS brane model. \\

\section*{Acknowledgments}
The work of D.W was supported through a 2016 PKPI Scholarship by the Directorate General of Resources for Science Technology and Higher Education of the Republic of Indonesia. T and JSK were supported by the Riset dan Inovasi Institut Teknologi Bandung and the Desentralisasi DIKTI research programs. \\


\end{document}